\documentclass[10pt]{iopart}

\usepackage{graphicx}               
\usepackage{psfrag}
\usepackage{dcolumn}                
\usepackage{bm}                     
\usepackage{cite}                     %
\usepackage{subfigure}

\newcommand{\ba}{\begin{array}}
\newcommand{\ea}{\end{array}}
\newcommand{\bc}{\begin{center}}
\newcommand{\ec}{\end{center}}

\newcommand{\eps}{\epsilon}

\newcommand{\UA}{\uparrow}
\newcommand{\DA}{\downarrow}

\newcommand{\phd}{^{\phantom{\dagger}}}
\newcommand{\SC}{{\cal S}}
\newcommand{\Dint}{\int {\cal D}[\SC]\;}
\newcommand{\YY}{{\cal Z}}

\newcommand{\JH}{{J_{\scriptsize\textrm{H}}}}
\newcommand{\Jeff}{{J_{\scriptsize\textrm{eff}}}}
\newcommand{\epol}{{\eps_{\scriptsize\textrm{pol}}}}
\newcommand{\ehole}{{\eps_{\scriptsize\textrm{hole}}}}
\newcommand{\tf}{{t_{\scriptsize\textrm{f}}}}

\begin{document}

\newcommand{\Eq}[1]{{Eq.~(\ref{#1})}}
\newcommand{\EQ}[1]{{Equation~(\ref{#1})}}
\newcommand{\av}[1]{{\left<{#1}\right>}}
\newcommand{\E}{{\textrm{e}}}

\newcommand{\cdag}{c^\dagger}
\newcommand{\cnod}{c^{\phantom{\dagger}}}
\newcommand{\adag}{a^\dagger}
\newcommand{\anod}{a^{\phantom{\dagger}}}
\newcommand{\ctdag}{\tilde c^\dagger}
\newcommand{\ctnod}{\tilde c^{\phantom{\dagger}}}

\title{Polaronic aspects of the two-dimensional ferromagnetic Kondo model} 

\author{M Daghofer\dag, W Koller\ddag, H G Evertz\dag, and W von der Linden\dag}
\address{\dag Institute for Theoretical and Computational Physics,
    Graz University of Technology, Petersgasse 16, A-8010 Graz,
    Austria.}
 \address{\ddag Department of Mathematics,
    Imperial College, 180 Queen's Gate, London SW7 2BZ, UK.}

\date{December 16, 2003}

\begin{abstract}
  The two-dimensional ferromagnetic Kondo model with classical corespins
  is studied via unbiased Monte Carlo simulations for a hole doping up to 
  $x=12.5\%$. A canonical
  algorithm for finite temperatures is developed.
  We show that with realistic parameters for the manganites
  and at low temperatures, the double-exchange mechanism does not lead
  to phase separation on a two-dimensional lattice but rather stabilises
  individual ferromagnetic polarons for this doping range.
  A detailed analysis of unbiased Monte Carlo results reveals that the
  polarons can be treated as independent particles for these hole
  concentrations.
  It is found that a simple polaron model describes the
  physics of the ferromagnetic Kondo model amazingly well.
  The ferromagnetic polaron picture provides an obvious explanation
  for the pseudogap in the one-particle spectral function
  $A_k(\omega)$ observed in the ferromagnetic Kondo model.
\end{abstract}


\ead{daghofer@itp.tu-graz.ac.at}

\section{Introduction}                                  \label{sec:intro}

Manganese oxides such as La$_{1-x}$Sr$_x$MnO$_3$,
La$_{1-x}$Ca$_x$MnO$_3$ and La$_{2-2x}$Sr$_{1+2x}$Mn$_2$O$_7$, which
have been thoroughly studied due to their colossal magnetoresistance
(CMR), show a very rich phase diagram depending on doping,
temperature, pressure and other parameters, see
e.g~\cite{proceedings98,Nagaev:book}. The phase diagram includes
ferromagnetic (FM), antiferromagnetic (AFM), paramagnetic (PM), charge-ordered,
metallic as well as insulating domains. The manganites  crystallise in the
perovskite like lattice structure, and quasi two-dimensional systems
with well separated  MnO$_2$-(bi)layers also exist. 

Crystal field splitting divides the five d-orbitals of the Mn ions into three
energetically favoured $t_{2g}$ and two $e_g$ orbitals. All three
$t_{2g}$ orbitals are singly occupied and rather localised. The
filling of the $e_g$ orbitals is determined by doping and these
electrons can hop from one Mn ion to the next via the intermediate
oxygen (double exchange, DE).
Due to a strong Hund's rule coupling, the spins of the three $t_{2g}$
electrons are aligned in parallel and form a corespin with
length $S =3/2$. Beinig localised, these electrons interact through
superexchange which leads to a weak antiferromagnetic coupling between
the corespins. Hund's rule coupling also leads to a
ferromagnetic interaction between the itinerant $e_g$ electrons
and the $t_{2g}$ corespin.

Apart from double- and superexchange, a complete description would have to
include Coulomb repulsion between the $e_g$ electrons, lattice
distortions and disorder.
Full quantum mechanical many-body calculations for a realistic model, including all
degrees of freedom, are not yet possible, see however \cite{GBB04pre}
for a one-dimensional study.
Several approximate studies of simplified models have therefore been
performed in order to unravel individual pieces of the rich phase
diagram of the manganites.
The electronic degrees of freedom are generally treated by a Kondo lattice
model~\cite{zener51}.

As full quantum mechanical results in more than one dimensions are
difficult to obtain for the quantum mechanical Kondo lattice model,
it has been  proposed to treat the S=3/2 corespins classically, see  
de Gennes~\cite{gennes60}, Dagotto {\it et al.}~\cite{dagotto98:_ferrom_kondo_model_mangan,
dagotto01:review} and Furukawa~\cite{furukawa98}
Unbiased Monte
Carlo (MC) techniques can then be applied. Coulomb
repulsion~\cite{hotta00:coo_ps_nn_coulomb}, classical phonons
\cite{yunoki98:_phase_separ_induc_orbit_degrees} and disorder
\cite{Motome_Furukawa_disorder} have
also been treated within this classical approach.
The validity of this approximation has been tested in
~\cite{EdwardsI,Nolting01,Nolting03,dagotto98:_ferrom_kondo_model_mangan} and
it appears that quantum effects are important for (S=1/2) corespins or at
$T=0$. For finite temperature and S=3/2, classical spins present a reasonable
approximation.

Further approximations can be made by taking into account, that the Hund
coupling $\JH$ is much stronger than the kinetic energy.
Consequently, configurations are very unlikely in which the
electronic spin is antiparallel to the local corespin. A customary approach
is to take $\JH \rightarrow \infty$. This approximation however
breaks down for the almost completely filled lower Kondo band. In the dilute hole
regime, the full Kondo model is governed by an effective AFM interaction
between the corespins due to excitations into the upper Kondo band. This
effect is completely absent from the $\JH \rightarrow \infty$
model.

An effective spinless fermion (ESF) model~\cite{KollerPruell2002a}
has been proposed to improve upon this approximation.
In this model, virtual excitations account for effects of
configurations, in which
the itinerant electron spin is antiparallel to the local corespin.
It has been demonstrated that the results of the ESF model are in excellent
agreement with those of the original Kondo model even for moderate values
of $\JH$.

For the FM Kondo with classical $t_{2g}$
corespins, elaborate Monte Carlo simulations have been performed in various dimensions~\cite{dagotto98:_ferrom_kondo_model_mangan,
yunoki98:_static_dynam_proper_ferrom_kondo,yunoki98:_phase,Yi_Hur_Yu:spinDE,dagotto01:review,furukawa98,
Motome_Furukawa_3dDE,Motome_Furukawa_Ucl,KollerPruell2002a,KollerPruell2002b,KollerPruell2002c,Aliaga_island_2d}
in order to determine  the physical properties of the DE model. For a review see for example
~\cite{dagotto01:review} and references therein. 
A two-dimensional Kondo lattice model for manganites in the $\JH \rightarrow
\infty$ limit has been thoroughly
investigated~\cite{Aliaga_island_2d} by means of MC calculations
similar to ours and
by analytical comparison of the groundstate energy for several phases.
Using a relatively high value for the antiferromagnetic exchange coupling,
Aliaga {\it et al.} find phase separation (PS), stripes, island phases
(small ferromagnetic domains that are stacked antiferromagnetically)
for commensurate fillings, and a so called ``Flux Phase''.
In a 2D-Kondo model applied to cuprates, stripes and a pseudogap are
observed in MC
simulations~\cite{Moraghebi_01:kondoCu,Moraghebi_02:kondoCu,Moraghebi_02b:kondoCu}. 

Many of these studies revealed features, e.~g. an infinite compressibility near the filled lower Kondo
band, which have been interpreted as signatures of PS.
PS has also been reported~\cite{Millis_PS} from computations based on a
dynamical mean field treatment based on the DE model at $T=0$. In previous MC
studies~\cite{KollerPruell2002a,KollerPruell2002b} for the  
DE model with classical core spins for 1D systems, we had obtained
numerical data comparable to those reported in Refs.~\cite{dagotto98:_ferrom_kondo_model_mangan,
yunoki98:_static_dynam_proper_ferrom_kondo,yunoki98:_phase}.
A detailed analysis of the data~\cite{KollerPruell2002c} revealed
however, that the aforementioned model with the standard parameter
set, relevant for the manganites, favours individual polarons over
phase separation. 
Other authors also found ferromagnetic polarons for the
almost empty lower Kondo band (i.~e. very few electrons) for S=1/2
corespins~\cite{Batista_FMPOL98,Batista_FMPOL00}, for the 1D AF Kondo Model with
few electrons~\cite{Homer_pol}, and for the 1D paramagnet at
higher temperatures~\cite{Aliaga_pol_1d}. Small
ferromagnetic droplets were predicted from energy considerations~\cite{Kagan_nano}.

In this paper, we present a numerical study of the
2D ferromagnetic Kondo model with classical corespins.
As in 1D, we find that the correct physical interpretation of the features which
have been interpreted as PS is rather given by ferromagnetic polarons,
i.e.\ small FM regions with {\em one single} trapped charge-carrier.
The polaron picture allows also a straight forward and obvious
explanation of the pseudogap, which has been previously observed in
the spectral density in experiments~\cite{DessauI, DessauII, DessauIII,Park95}
and MC simulations~\cite{dagotto01:review,KollerPruell2002a}.  
Experimental evidence for small FM droplets in low doped
La$_{1-x}$Ca$_{x}$MnO$_3$ has been reported~\cite{Biotteau_01,Hennion_98}.

This paper is organised as follows.
In Sec.~\ref{sec:model} the model Hamiltonian is presented and
particularities of the MC simulation for the present model are outlined. A
canonical algorithm is introduced.
In Sec.~\ref{sec:FMP}, we introduce a simplified model of ferromagnetic
polarons embedded in an AFM background and present results for this
model. In Sec.~\ref{sec:Numerical_Results}, these results are compared to unbiased Monte Carlo simulations
of the 2D ferromagnetic Kondo model with classical corespins at realistic
parameter values. The key results of the
paper are summarised in Sec.~\ref{sec:conclusion}. 

\section{Model Hamiltonian and unbiased Monte Carlo}        \label{sec:model}

In this paper, we will concentrate solely on properties of the
itinerant $e_g$ electrons interacting with the local $t_{2g}$ corespins.
We also neglect the degeneracy of the $e_g$ orbitals.
The degrees of freedom of the $e_g$ electrons are then described by a
single-orbital Kondo lattice model~\cite{KollerPruell2002b}.
As the corespins are approximated by classical spins, they are replaced by unit vectors
$\mathbf S_i$, parameterised by polar and azimuthal angles  $\theta_i$ and
$\phi_i$, respectively.
The magnitude of both corespins and $e_g$-spins is absorbed into the exchange
couplings.

\subsection{Effective Spinless Fermions (ESF)}                   \label{subsec:ESF}

By choosing the quantisation of the $e_g$ spin parallel to the local
$t_{2g}$ corespin, simplified low-energy models for fillings $0 \leq
N_{\scriptsize\textrm{el}} \leq 1$ (i.e. for the lower Kondo band) can be derived,
namely the $\JH \to \infty$ approximation and the effective spinless
fermion model with finite $\JH$ (see~\cite{KollerPruell2002a}):
\begin{equation}                                       \label{eq:H}
  \hat H = -\sum_{<i,j>} t^{\UA\UA}_{i,j}\,
    \cdag_{i}\,\cnod_{j} - \sum_{i,j}
    \frac{t^{\UA\DA}_{i,j}\,t^{\DA\UA}_{j,i}}{2\JH}\, \cdag_{i}\cnod_{i}
    + J'\sum_{<i,j>} \mathbf S_i \cdot \mathbf S_j \;.
\end{equation}
The spinless fermion operators~$\cnod_{j}$ correspond to \emph{local} spin-up electrons
(i.e. parallel to the corespin) only.
The spin index has therefore been omitted.
With respect to a {\em global} spin-quantisation axis the ESF
model~(\ref{eq:H}) still contains contributions from both spin-up and
spin-down electrons.

The first term in \Eq{eq:H} corresponds to the kinetic energy in
tight-binding approximation.
The modified hopping integrals $t^{\sigma,\sigma'}_{i,j}$
depend upon the $t_{2g}$ corespin orientation
\begin{equation}                                      \label{eq:modihop}
  t^{\sigma,\sigma'}_{i,j} \;=\;
  t_{0}\; u_{i,j}^{\sigma,\sigma'}\;,
\end{equation}
where the relative orientation of the $t_{2g}$ corespins at site $i$ and
$j$, expressed by the angles $0\leq\vartheta\leq\pi$ and
$0\leq\phi<2\pi$, enters via
\begin{equation}
  \begin{array}{ccc}
    u^{\sigma,\sigma}_{i,j}(\SC)  =&
    c_ic_j + s_is_j\E^{i(\phi_j - \phi_i)}&=
    \cos(\vartheta_{ij}/2)\;\E^{i\psi_{ij}} \\
    u^{\sigma,-\sigma}_{i,j}(\SC) =&
    c_is_j\E^{-i\phi_j} + s_ic_j\E^{-i\phi_i}&=
    \sin(\vartheta_{ij}/2)\;\E^{i\chi_{ij}}
  \end{array}
\end{equation}
with the abbreviations $c_i=\cos(\vartheta_i/2)$ and $s_i=\sin(\vartheta_i/2)$.
These factors depend on the relative angle $\vartheta_{ij}$ of
corespins $\mathbf S_i$ and $\mathbf S_j$ and on some complex phases
$\psi_{ij}$ and $\chi_{ij}$. Because hopping is largest for parallel
corespins, this term favours ferromagnetism.
For certain spin structures, an electron may obtain
a different phase depending on the path taken from one lattice
site to another. An example of such structures is the so called Flux
phase~\cite{Aliaga_island_2d,Agterberg_00,Yamanaka_98}.

The second term in \Eq{eq:H} accounts for virtual hopping processes to
antiparallel spin--corespin configurations and vanishes in the limit
$\JH \to \infty$. For finite $\JH$, the ESF model takes
into account virtual hopping processes to antiparallel spin-corespin
configurations much in the same way as the $tJ$-model includes virtual
hopping to doubly occupied sites for the Hubbard model.
Because this term is proportional to the density, it is
most relevant near the completely filled lower Kondo band, where the
kinetic energy is moreover reduced.
The last term is a small antiferromagnetic exchange of the corespins.

The hopping strength $t_0$ will serve as our unit of energy.
$\JH$ is usually taken to be of the order of magnitude
of $4 t_0$ to $8 t_0$, and $J'$ of the order of $t_0 / 100$.
\subsection{Grand Canonical Treatment}                 \label{subsec:GrCanAlg}

We define the grand canonical partition function as
\begin{equation}             \label{eq:Y}
  \begin{array}{ccc}
    \YY(\mu) &=& \Dint \textrm{ tr}_c\, \E^{-\beta (\hat H(\SC)-\mu \hat{N})}\\[1ex]
    \Dint &=&  \prod_{i=1}^L\;\Big(\int_{0}^{\pi} d\theta_i\sin \theta_i
    \int_{0}^{2\pi} d\phi_i\Big)\;,
  \end{array}
\end{equation}
where $\textrm{ tr}_c$ indicates the trace over fermionic degrees of freedom at
inverse temperature $\beta$ and chemical potential $\mu$, and $\hat{N}$ is
the operator for the total number of $e_g$ electrons.  As $\hat H$ is
a one-particle Hamiltonian, the fermionic trace can easily be carried
out using the free fermion formula, yielding the statistical weight of
a corespin configuration $\SC$
\begin{equation}                                  \label{eq:MC_weight}
    w(\SC|\mu) = \frac{\textrm{ tr}_c\, \E^{-\beta (\hat H(\SC)-\mu \hat{N})}}{\YY(\mu)}\;.
\end{equation}
\EQ{eq:Y} is the starting point of grand canonical Monte Carlo simulations of the Kondo
model~\cite{dagotto98:_ferrom_kondo_model_mangan}
where the sum over the classical spins is performed via Markov chain importance sampling;
the spin configurations $\SC$ are sampled with the probability determined by the
weight factor $w(\SC|\mu)$.

In order to avoid the CPU expensive full diagonalization of the
one-particle Hamiltonian, Motome and
Furukawa~\cite{Motome_Furukawa_kpm} suggested to replace it with an
expansion in Chebychev polynomials. The CPU time then
scales with the system size $L$ as $O(L^2\log(L))$ instead of $O(L^3)$ as for
the full diagonalization and the algorithm can be easily parallelised.
We found, however, that on single processors the full diagonalization can 
be accelerated to be faster than this approach up to system sizes of
$10^3$ lattice sites, while both algorithms would be too slow
for the study of larger systems on present day's processors. The key for a faster full
diagonalization is to exploit the structure of the Hamiltonian: The lattice sites are
relabeled in order to obtain a band matrix with as few diagonals as
possible. This is only an alternative assignment of the linear index 
to the two dimensional lattice vector and does therefore not introduce any
approximation or error. Fast library routines for band matrices can then be used.
Since the purpose of our MC code was to perform parameter studies, we
did not program a parallel algorithm but instead ran the whole
calculation on each CPU with a different parameter set.
Recently, an $O(L)$ algorithm has been proposed by the same authors,
which reduces the numerical effort by approximating the matrix-vector
multiplication~\cite{Motome_Furukawa_Ucl}.

In the 2D case we have employed MC updates in which single spins were
rotated.
The angle of rotation was optimised to keep the acceptance high enough.
From time to time a complete flip $\mathbf S_i \to -\mathbf S_i$ was
proposed. The skip between subsequent measurements was chosen to be 50 to a few hundreds of
lattice sweeps reducing autocorrelations to a negligible level.
We have performed MC runs with some hundreds to 2000 measurements on a $12 \times 14$ lattice.
This geometry was chosen to reduce finite size (closed shell) effects
observed on a square lattice. The number of measurements was higher for
calculations in the polaronic regime, where the particle
number fluctuates strongly, in order to have sufficient measurements for each filling.

As previously shown~\cite{KollerPruell2002b}, the spin-integrated one-particle
Green's function in global quantisation can be written as
\begin{equation}                                          \label{eq:MC_GF2}
  \sum_\sigma
  \ll a\phd_{i \sigma}; a^\dagger_{j \sigma} \gg_\omega
  = \Dint w(\SC|\mu) u_{ji}^{\UA\UA}(\SC)
    \ll c\phd_{i}; c^\dagger_{j} \gg^\SC_\omega\;,
\end{equation}
where $\ll c\phd_{i}; c^\dagger_{j} \gg^\SC_\omega$ is the
Green's function in local spin quantisation.
It can be expressed in terms of the
one-particle eigenvalues $\eps^{(\lambda)}$ and the corresponding
eigenvectors $\psi^{(\lambda)}$ of the Hamiltonian $\hat H(\SC)$:
\[
\ll \cnod_{i}; \cdag_{j}\gg^\SC_\omega
= \sum_\lambda \; \frac{\psi^{(\lambda)}(i) \;\psi^{*(\lambda)}(j)}
{\omega - (\eps^{(\lambda)} -\mu) + i0^+ }
\]
It should be pointed out that the one-particle density of states (DOS)
is identical in global and local quantisation; for details see~\cite{KollerPruell2002b}.

\subsection{Canonical Algorithm}              \label{subsec:CanAlg}

Due to the jump in the electron density at the critical chemical
potential (shown later in Fig.~\ref{n_vs_mu_JH06J002beta50.eps}), some
electron fillings cannot be examined with the grand canonical
algorithm.
We therefore developed a canonical scheme.
Canonical calculations were done by computing the eigenenergies for
each corespin configuration and then filling the available electrons
into the lowest levels~\cite{Aliaga_island_2d}. 
This method does, however, not account for thermal particle-hole excitations
around the Fermi energy. When several one-particle states have similar
energy, these may become important. It was also proposed to adjust the chemical
potential by solving an implicit equation using the
Newton-Raphson algorithm for every single spin configuration to give the
wanted particle number~\cite{aliaga_dagotto_ce_can}.
While this approach includes particle hole excitations, it is still
not certain, whether the calculation is correct and small differences
possibly have a considerable effect when phases with and without a
pseudogap compete.

On the other hand, an exact approach would mean calculating the Boltzmann
weight for every possible distribution of $N_{el}$ particles on $L$
energy levels and summing over their contributions.
Even for small lattice sizes $L$, this clearly becomes too demanding
for more than a few electrons or holes.
Instead, we took into account just the lowest excitations of the Fermi sea
by filling $N_{el}^0<N_{el}$ electrons into the $N_{el}^0$ lowest states and
considering only the distributions of the $N_{el} - N_{el}^0$ remaining
electrons on the states around the Fermi energy. Usually, it is sufficient to
take $N_{el} - N_{el}^0 \approx 5$.
The weight for the corespin configuration $\SC$ then depends on the particle
number instead of the chemical potential:
\begin{equation}  \label{eq:MC_weight_can}
      w(\SC|N_{el}) = \frac{\sum_{\tilde{\mathcal{P}}}\, \E^{-\beta \hat
      H(\SC,\tilde{\mathcal{P}}(N_{el}) )}}{\YY(N_{el})}\;,
\end{equation}
where $\tilde{\mathcal{P}}$ denotes these restricted permutations.

Although this is more time-consuming than the grand canonical calculation
of the fermionic weight, the additional consumption of computer time is small
compared to the time needed for the diagonalization of the one-particle
Hamiltonian. 
The particle-hole excitations, which are thus included, can be crucial when
examining competition between phases with and without a pseudogap.

A MC update - especially a complete spin flip - may lead to a configuration which is very unlikely to
occur at the given particle number, although it may be a good configuration
for a different filling. A later MC move might then lead back to the original
particle number, and these moves should improve autocorrelation. We therefore
allow density fluctuations within a set of four
to five particle numbers.
In order to spend a comparable number of MC steps at each filling, prior weight
factors $g(N_{el})$ were introduced and adjusted in a prerun, giving
\begin{equation}                                  \label{eq:MC_weight_sw_can}
    w(\SC) = \sum_{N_{el}=N_{min}}^{N_{max}} w(\SC | N_{el})\ g(N_{el})\;.
\end{equation}
The sum is taken over the set of allowed particle numbers $N_{min} \leq
N_{el} \leq N_{max}$ and $w(\SC |
N_{el})$ is calculated according to \Eq{eq:MC_weight_can}.

When evaluating observables for fixed electron number, one has to calculate
the expectation value
\begin{equation}                                  
    \langle\;\mathcal{O}\;\rangle_{N_{el}} =
    \frac{\sum_{\SC} \mathcal{O}(\SC)_{N_{el}}\
       w(\SC | N_{el})}
    {\sum_{\SC} w(\SC | N_{el})}\;,
\end{equation}
which can be rewritten as
\begin{equation}                                  
    \langle\;\mathcal{O}\;\rangle_{N_{el}} =
    \frac{\sum_{\SC} \mathcal{O}(\SC)_{N_{el}}
      \frac{w(\SC | N_{el})}{w(\SC)}\ w(\SC)}
    {\sum_{\SC} \frac{w(\SC | N_{el})}{w(\SC)} w(\SC)}\;,
\end{equation}
Configurations $\SC$ occur in the Markov chain with probability proportional
to $w(\SC)$;
when the sum is taken over the configurations produced by the MC run, the
expectation value therefore becomes
\begin{equation}                                  \label{eq:obs_sw_can}
  \langle\;\mathcal{O}\;\rangle_{N_{el},\;MC} = 
  \frac{\sum_{\SC,MC} \mathcal{O}(\SC)_{N_{el}}
      \frac{w(\SC | N_{el})}{w(\SC)}}
  {\sum_{\SC,MC} \frac{w(\SC | N_{el})}{w(\SC)}}\;.
\end{equation}

\section{Ferromagnetic Polaron Model}                              \label{sec:FMP}

Near half filling of a single $e_g$ band, a tendency toward
phase separation has been reported in various computational studies~\cite{Millis_PS,dagotto98:_ferrom_kondo_model_mangan,
yunoki98:_static_dynam_proper_ferrom_kondo,yunoki98:_phase}.

In most cases, the existence of phase separation is inferred from
a discontinuity of the electron density as a function of the chemical
potential.
At the critical chemical potential where this discontinuity is found,
it is claimed that the system separates into FM domains of high carrier
concentration and AFM domains of low carrier concentration.

We have already shown that for a 1D system~\cite{KollerPruell2002c}
this picture is in general incorrect.
In fact, what happens is that each single hole is dressed by
a ferromagnetic cloud in which it delocalizes. The system can be well
described by free quasiparticles consisting of a single hole plus a local
(three to four-sites for 1D, 5 sites for 2D) ferromagnetic well embedded in an AF background.
Each of these added quasiparticles gains the same energy, which is exactly
balanced by the energy to be paid for the critical chemical potential $\mu^*$. Hence the discontinuity
of the particle number at low temperatures.

\begin{figure}
  \centering
  \subfigure[]{\includegraphics[width=0.46\textwidth]
    {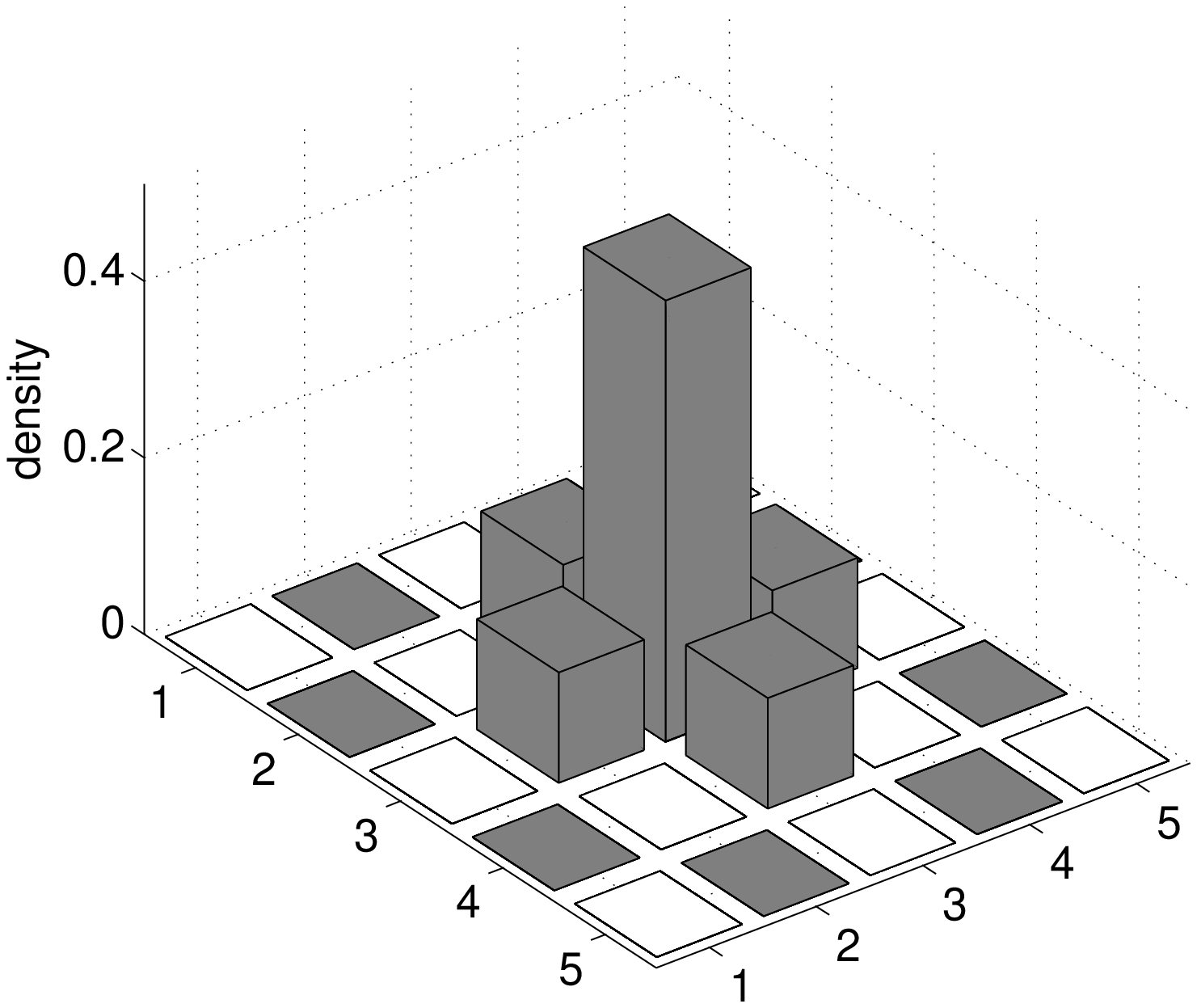}\label{polaron_conf}}\hfill 
  \subfigure[]{\includegraphics[width=0.46\textwidth]
    {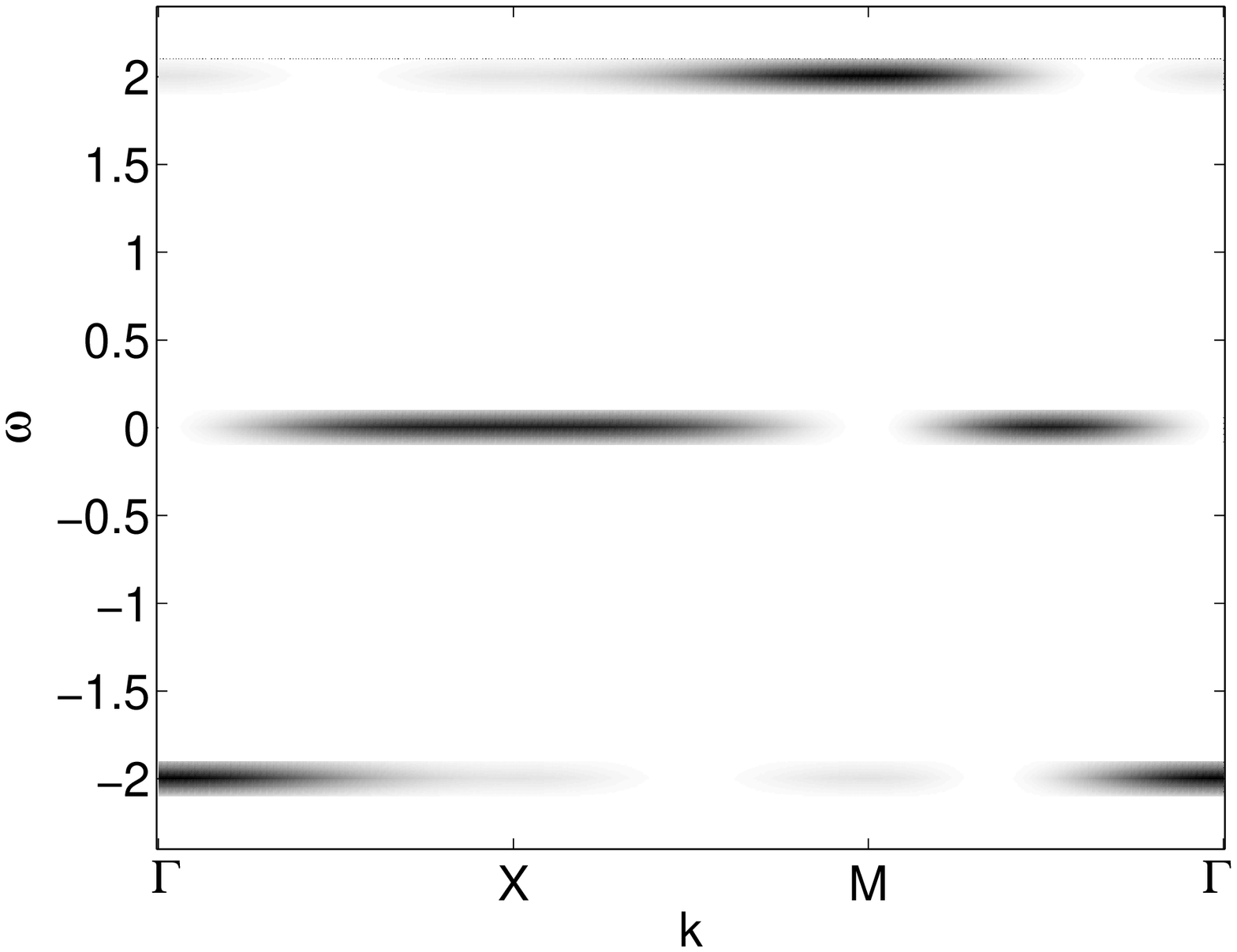}\label{polaron_bands}}\\
  \caption{Idealised FM polaron of $L_{\scriptsize\textrm{f}}=5$ lattice sites,
  embedded in an AFM background. 
  (a) Spin and hole-density configuration for the ground state. Empty
  (filled) squares represent spin down (up). Height represents hole
  density.
  (b) Contribution of the polaron to the one-particle spectral
  function. For visibility, the $\delta$-peaks in the spectral density 
  have been broadened to a width of $0.2$.}
  \label{polaron.eps}
\end{figure}

Here we show that ferromagnetic polarons, i.e.\ {\em single} charge carriers
surrounded by small ferromagnetic spin-clouds, are indeed formed when holes are
doped into a completely filled lower Kondo band in 2D.
In this section we discuss the properties of idealised two dimensional model polarons,
whereas in the next section we compare our polaron-model to unbiased Monte
Carlo results.
\subsection{One single polaron}
As reference configuration we consider the completely filled lower Kondo
band where super exchange and the contribution from the virtual hopping
process, see \Eq{eq:H},  give rise to an antiferromagnetic corespin pattern.
The smallest defect in a completely antiferromagnetically ordered 2D lattice
is the flipping of one single spin. Then a five-site ferromagnetic region
forms.
When we introduce a hole into the system, it can delocalize in this
region because of the double-exchange mechanism. As there is no
double-exchange hopping between sites with perfectly antiferromagnetic
spins, the hole is trapped inside the ferromagnetic region.
In this simple model, the hole can hop from the central site of this region
(site 1) where the spin has been flipped, to all nearest neighbours (sites 2--5).
The tight-binding Hamiltonian describing this situation reads
\[
  H_{\scriptsize \textrm{hole}} = -\tf \,\left(\begin{array}{ccccc}
      0 & 1 & 1 & 1 & 1\\
      1 & 0 & 0 & 0 & 0\\
      1 & 0 & 0 & 0 & 0\\
      1 & 0 & 0 & 0 & 0\\
      1 & 0 & 0 & 0 & 0 \end{array}\right)\;,
\]
with the hopping between ferromagnetic sites $\tf$ in this model.
Since the Hamiltonian is symmetric with respect to rotations by $\pi/2$, the
ground state should also show this property and can thus be found in the
space spanned by $(1,0,0,0,0), (0,1/2,1/2,1/2,1/2)$, where the Hamiltonian
reads
\[
  H_{\scriptsize\textrm{hole}} = -\tf \,\left(\begin{array}{cc}
      0 & 2 \\ 2 & 0 \end{array}\right)\,.
\]
The delocalization energy of the hole is thus given by
$\ehole=-2\,\tf$. This energy can
be gained as a hole delocalizes in the ferromagnetic domain. The
ground state $1/\sqrt{2}\;(1,1/2,1/2,1/2,1/2)$ has the hole density depicted in 
Fig.~\ref{polaron_conf}. Excited states are found at $\eps=+2\,\tf$ and
$\eps=0$. The highest state at $\eps=+2\,\tf$ is given by
$1/\sqrt{2}\;(-1,1/2,1/2,1/2,1/2)$ and does also have $s$-symmetry, while the
states at $\eps=0$ have $p$- and $d$-symmetry. They appear in the
one-particle spectral function of the configuration shown in
Fig.~\ref{polaron_bands}.

To create such a ferromagnetic domain, however, four antiferromagnetic bonds
have to be broken. This costs the energy of $2\times 4\, \Jeff$.
Near the completely filled lower Kondo band $x=0$, the total
antiferromagnetic exchange coupling is approximately given by $\Jeff =
1/(2 \JH) + J'$, see~\cite{KollerPruell2002c}.
The energy gained by adding one hole to the system thus reads
\begin{equation}
  \epol = -2\,\tf + 8  \Jeff\label{eq:e_pol}
\end{equation}
When the chemical potential approaches $-\epol$ from above, holes
start to enter the system forming individual polarons. Therefore, the
critical chemical potential is given by $\mu^* = -\epol$.

Up to a certain concentration, these holes can be treated as free fermions
which all have the same energy $\epol$.
The energy may, however, depend on the temperature if $\tf$ or
$\Jeff$ do.
The more obvious temperature effect is the smearing of the discontinuity in
the electron filling at higher temperatures, which results from the
application of the Fermi-Dirac statistics to these quasiparticles.

\begin{figure}
  \centering
  \subfigure[]{\includegraphics[width=0.46\textwidth]
    {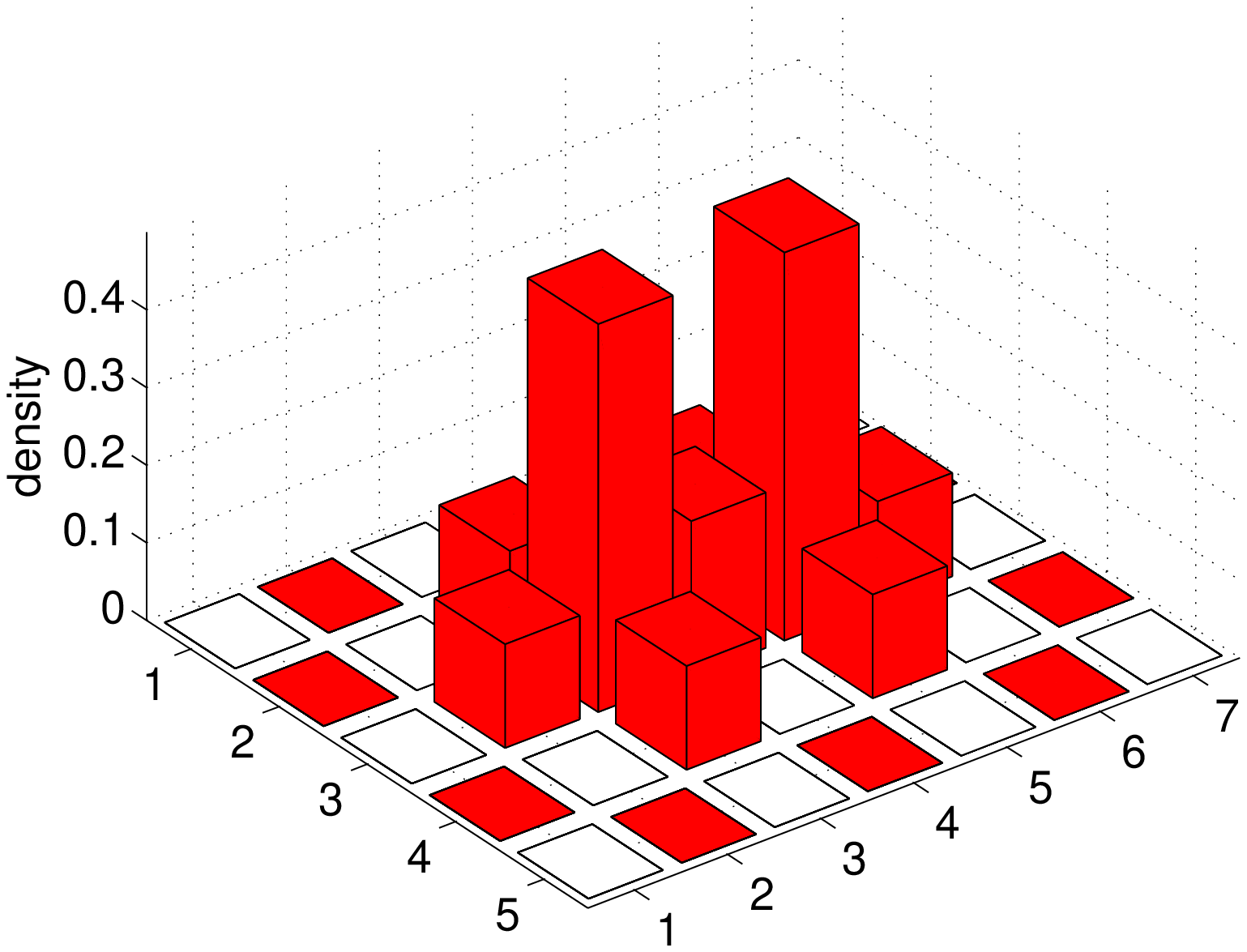}\label{dbl_polaron_1}}\hfill 
  \subfigure[]{\includegraphics[width=0.46\textwidth]
    {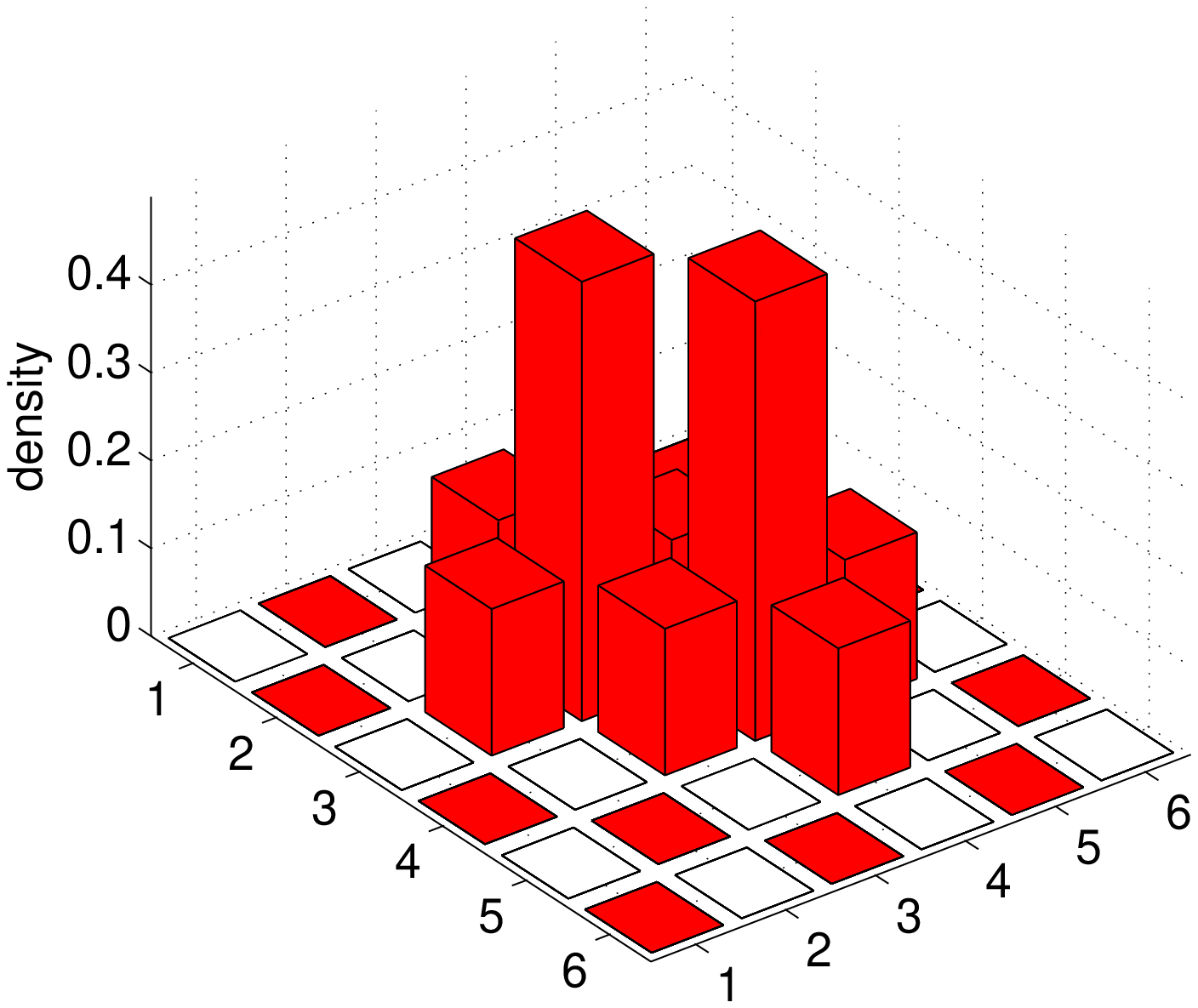}\label{dbl_polaron_2}}\\
  \caption{Idealised pictures of possible spin configuration containing two
  holes. Empty (filled) squares represent spin down (up). Height represents hole
  density.}  \label{dbl_polaron}
\end{figure}

It is interesting to compare the energy of two such independent holes to a
larger FM formation containing two holes. The two possibilities with the
least broken AFM bonds are depicted in
Fig.~\ref{dbl_polaron}. For both configurations, eight AFM bonds have to be
broken, the same as for two polarons. For the formation in Fig.~\ref{dbl_polaron_1},
the sum of the two lowest eigenenergies is however
$\epsilon_1+\epsilon_2=-3.97\,\tf$, i.\,e. the kinetic energy gained is smaller than for
two independent polarons, where it is $2 \ehole = -4\,\tf$. For the
second formation Fig.~\ref{dbl_polaron_2}, the energy gain is even lower
($\epsilon_1+\epsilon_2=-3.86\,\tf$). The larger FM structures with
two holes therefore have higher energy than two separate holes and are
energetically disfavoured, although the difference is not very large,
especially for the structure depicted in Fig.~\ref{dbl_polaron_1}. 

\subsection{Results for a small number of polarons}
\begin{figure}
  \centering
  \includegraphics[width=0.33\textwidth,height=0.2\textheight]
                  {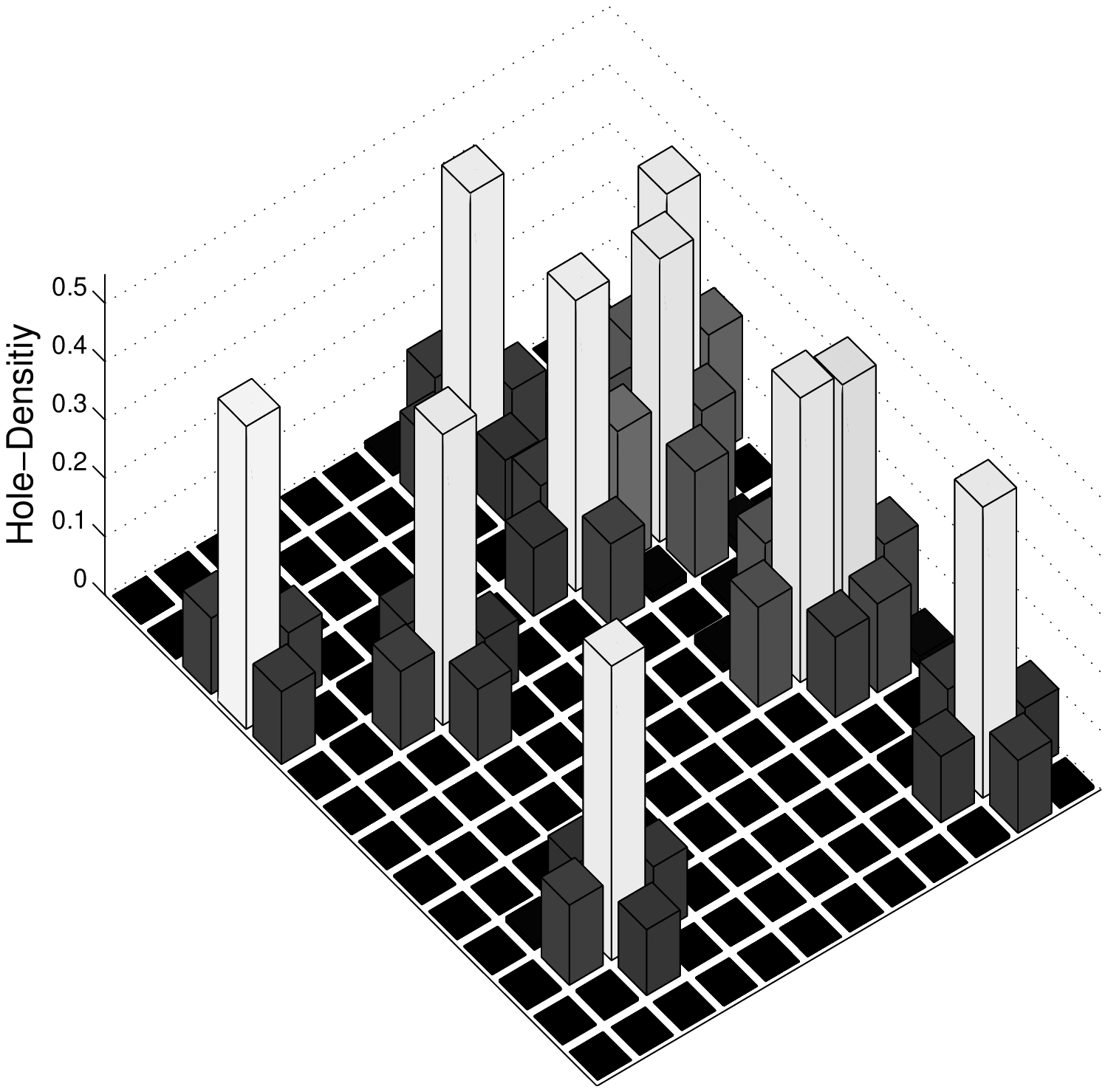}
  \includegraphics[width=0.33\textwidth,height=0.2\textheight]
                  {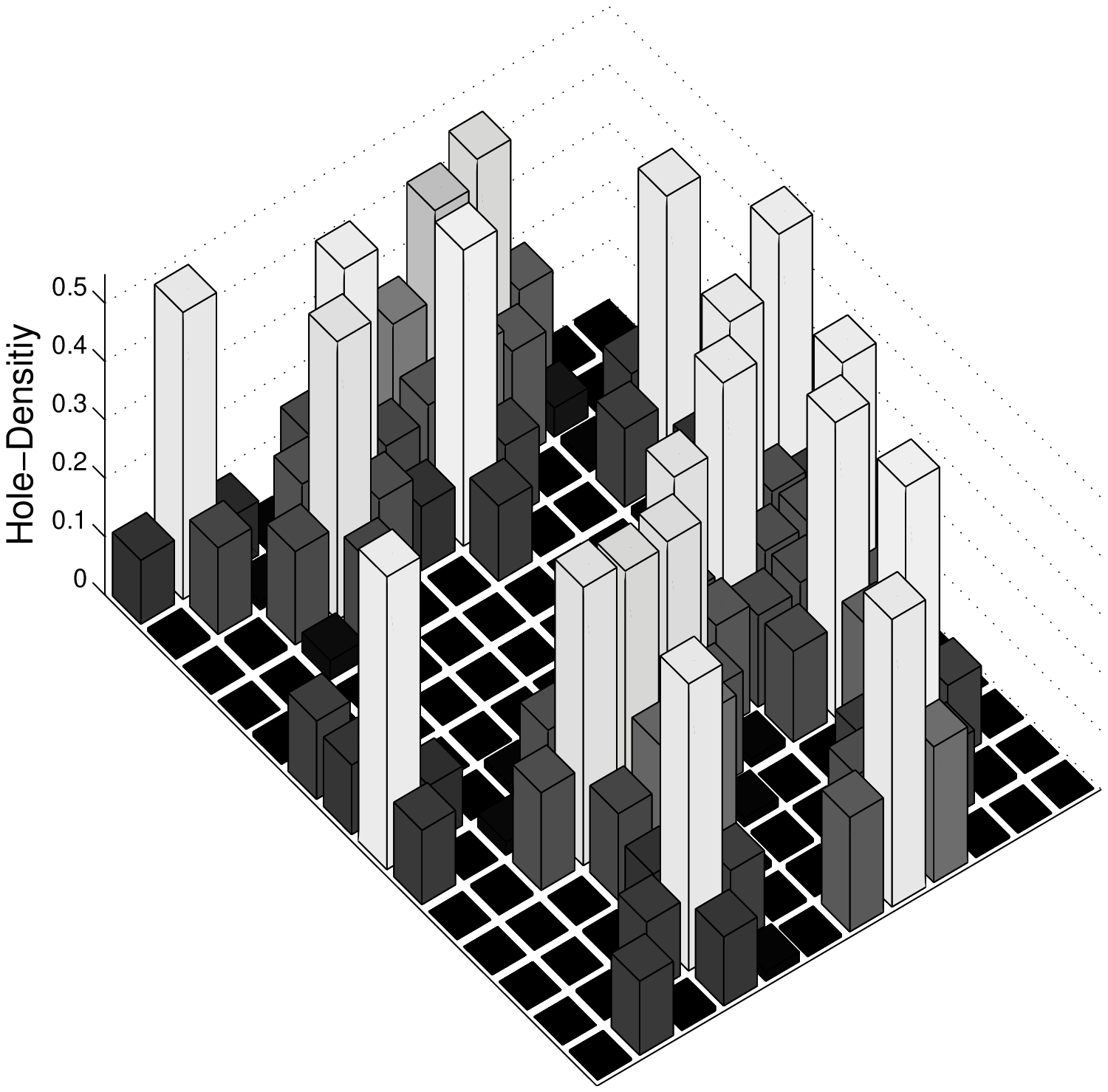}
  \caption{Representative hole-density configuration for $10$ (left, $x\approx 0.06$)
    and $20$ (right, $x\approx 0.12$) holes in the simplified polaron model on a
    $12\times 14$ lattice. The holes are located
    where one spin is flipped from the AFM reference
    configuration. The results correspond to $\beta=50, J'=0.02,
    \JH=6$. Grey shades are for better visibility.}
  \label{fig:Polsnap}
\end{figure}

\begin{figure}
  \centering
  \includegraphics[width=0.6\textwidth,trim= 0 50 0 50,clip]
                  {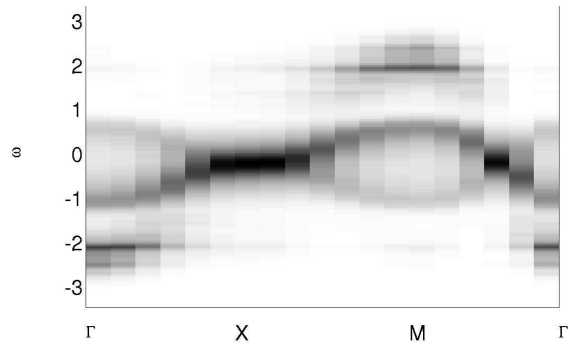}
  \caption{Spectral density of the polaron model for $20$ holes on a
    $12\times 14$ lattice ($x\approx 0.12$), using the same
    parameter values as Fig.~\ref{fig:Polsnap}.} 
  \label{polaron20_bands.eps}
\end{figure}

In this subsection we push our polaron ideas further to treat the case
of a small number of holes in an AFM background. To this end, we perform
a simple simulation.
We start with a perfect AFM reference configuration.
We then add $N_{\scriptsize\textrm{holes}}$ holes by
flipping $N_{\scriptsize\textrm{holes}}$ randomly chosen spins, excluding flipping a
spin back. Finally, we add random deviations to each corespin
in order to account for thermodynamic fluctuations. These fluctuations lead to
a finite DE hopping amplitude in the AFM band and their size is therefore
fitted to match the bandwidth observed in the
MC results, see Sec. \ref{sec:Numerical_Results}. We then
diagonalise the resulting ESF-Hamiltonian and compute observables in the
canonical ensemble with $N_{\scriptsize\textrm{el}} =
 N_{\scriptsize\textrm{sites}} - N_{\scriptsize\textrm{holes}} =
 N_{\scriptsize\textrm{sites}} - N_{\scriptsize\textrm{flipped spins}}$
 as explained in Sec.~\ref{subsec:CanAlg}.
The observables are averaged over many such configurations. 
Typical hole-density configurations are depicted in
Fig.~\ref{fig:Polsnap} for $10$ and $20$ holes, i.~e. $10$ resp. $20$
spins were flipped from the initial perfect AF
configuration. In the formulae for the ESF-Hamiltonian and the observables,
the parameters were set to $\beta=50, J'=0.02$ and $\JH=6$. 
It should be emphasised that the probability distribution used in
choosing the spins to be flipped is completely flat.
It is only ensured that exactly $N_{\scriptsize\textrm{holes}}$ spins are flipped from
the AFM reference configuration. The chosen parameter values therefore do not influence
the obtained spin configurations. 

As an example for observables calculated in this pure polaron model,
we show the one-particle spectral function for $20$ holes in
Fig.~\ref{polaron20_bands.eps}. The center is occupied by the AFM
tight-binding band with a mirror band due to the
doubling of the unit cell in the AFM spin configuration.
At $\omega=\pm 2$, the polaronic states can be clearly seen.
In addition to them, one sees a number of weaker signals in the
vicinity of $\omega=\pm 2$. These
stem from larger ferromagnetic regions, i.~e. from contiguous polarons. 

\section{Unbiased MC Results}                       \label{sec:Numerical_Results}

In this section we present results of unbiased Monte Carlo simulations for
$J' = 0.02$ and $\JH=6$ and show that they correspond to independent
ferromagnetic polarons.

\begin{figure}
  \centering
  \includegraphics[width=0.5\textwidth]{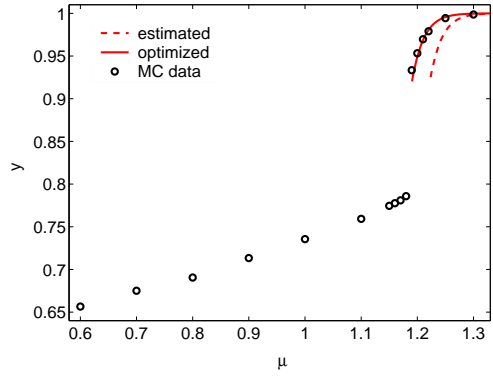}
  \caption{Electron density $y$ as a function of the chemical potential $\mu$
    of a $14\times 12$ lattice for $\beta=50, J'=0.02$ and $\JH=6$.
    Circles indicate MC results, the solid (dashed) line shows the free fermion
    results for the fitted (estimated) energy.}
  \label{n_vs_mu_JH06J002beta50.eps}
\end{figure}

Figure \ref{n_vs_mu_JH06J002beta50.eps} shows the electron density as a
function of the chemical potential at $\beta = 50$.
Depending on the value of the hopping parameter $t_0$, this is in
a range of $50K - 100K$ , i.~e. relevant for experiments. There is a
discontinuity in the density (infinite compressibility), but one observes
that the electron number does not drop at once from 1 (AFM) to
0.8 (FM). Instead, it first
decreases only slowly from the completely filled band, the slope of the curve then
becomes gradually steeper until it is vertical.
For a qualitative description of the MC results by the polaron model, we use
$\tf=t_0$ and $\Jeff= 1/(2 \JH)+J'$ 
which yields a polaron energy of $\epol \simeq -1.17$.
Using this value for the critical chemical potential we obtain the dashed
line. Although it is shifted by some constant energy, it already correctly
reflects the trend in the Monte Carlo data.
Much better agreement can be found by fitting the polaron energy to the
Monte Carlo data. In our case we obtain $\epol \simeq -1.14$.
The corresponding Fermi function is shown as the solid line.

\begin{figure}
  \centering
  \includegraphics[height=0.33\textwidth]{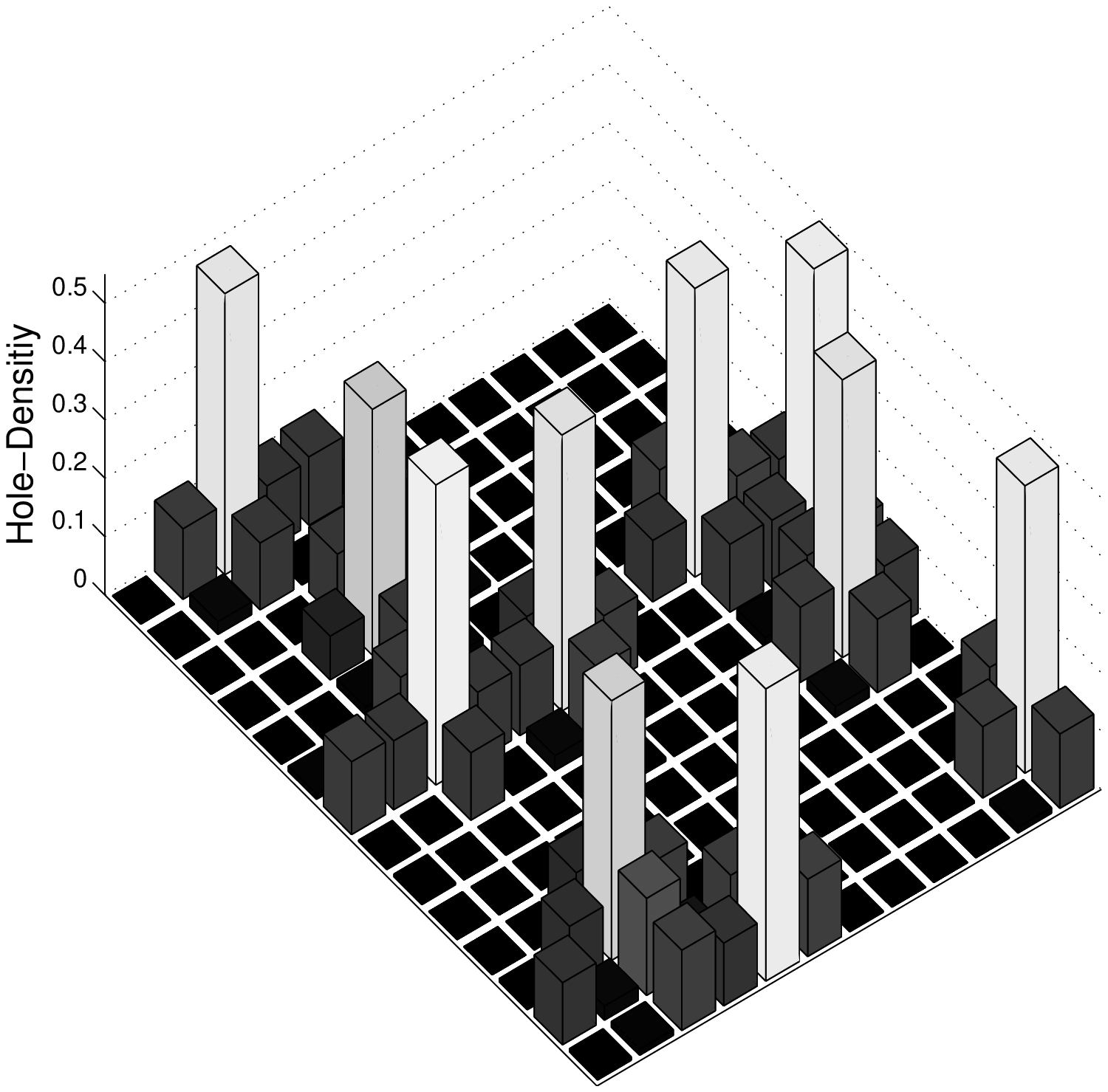}
  \includegraphics[height=0.33\textwidth]{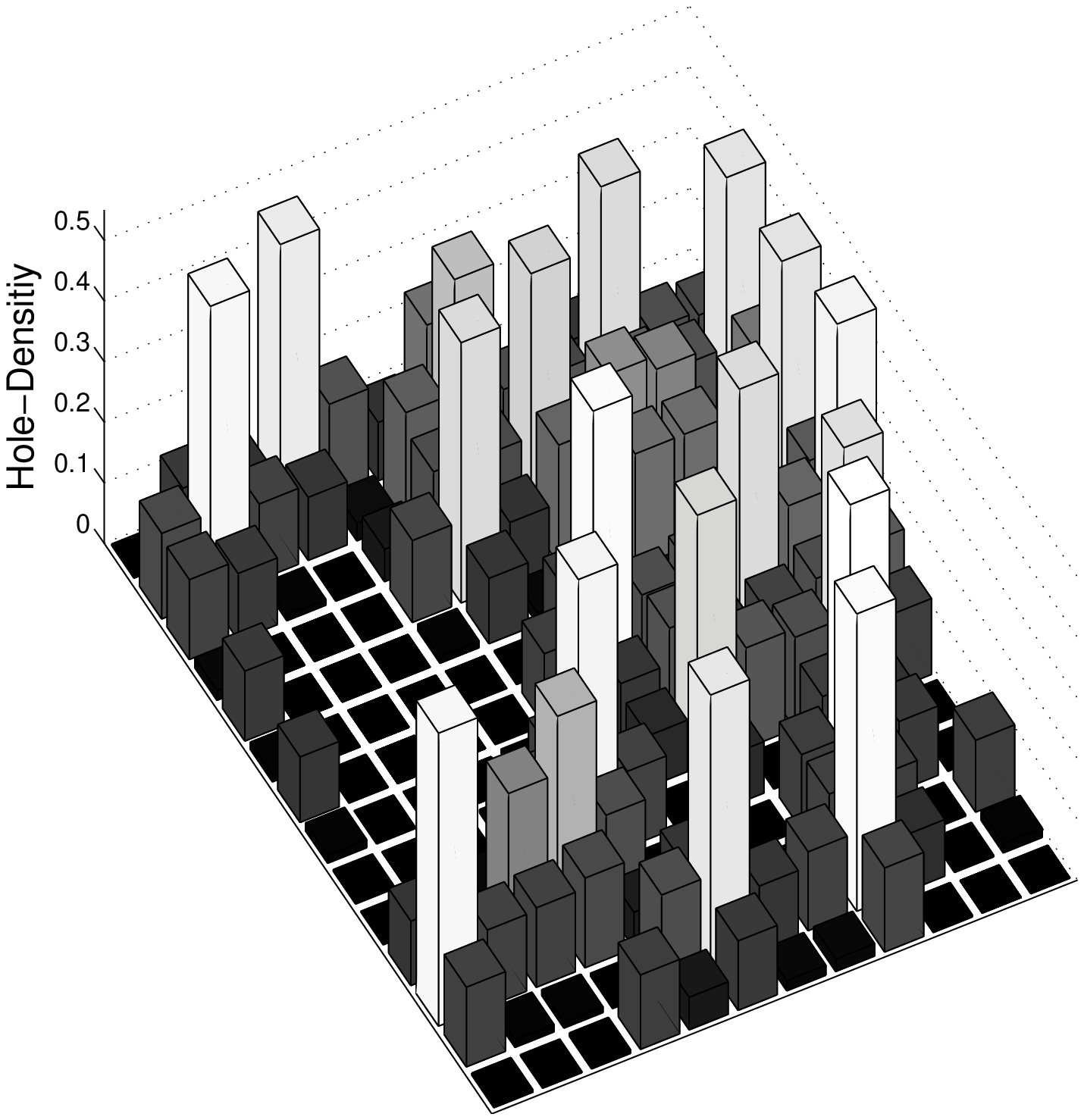}
  \caption{MC snapshot of the hole density for $10$ (left) resp.~$20$
    (right) holes in a $14\times 12$ lattice at $\beta=50, J'=0.02, \JH=6$. }
  \label{MC_snapshot_J0.02_beta80_148.eps}
\end{figure}

Figure~\ref{MC_snapshot_J0.02_beta80_148.eps} shows MC snapshots with $10$ and
$20$ holes.The polarons can be clearly seen, $10$
polarons for $10$ holes and $19$ polarons for $20$ holes. Only the $20^{th}$
hole at the larger doping is delocalised. There is an obvious similarity to
the idealised polaron model, see Fig.~\ref{fig:Polsnap}.

\begin{figure}
  \centering
  \subfigure[]{\includegraphics[width=0.4\textwidth]{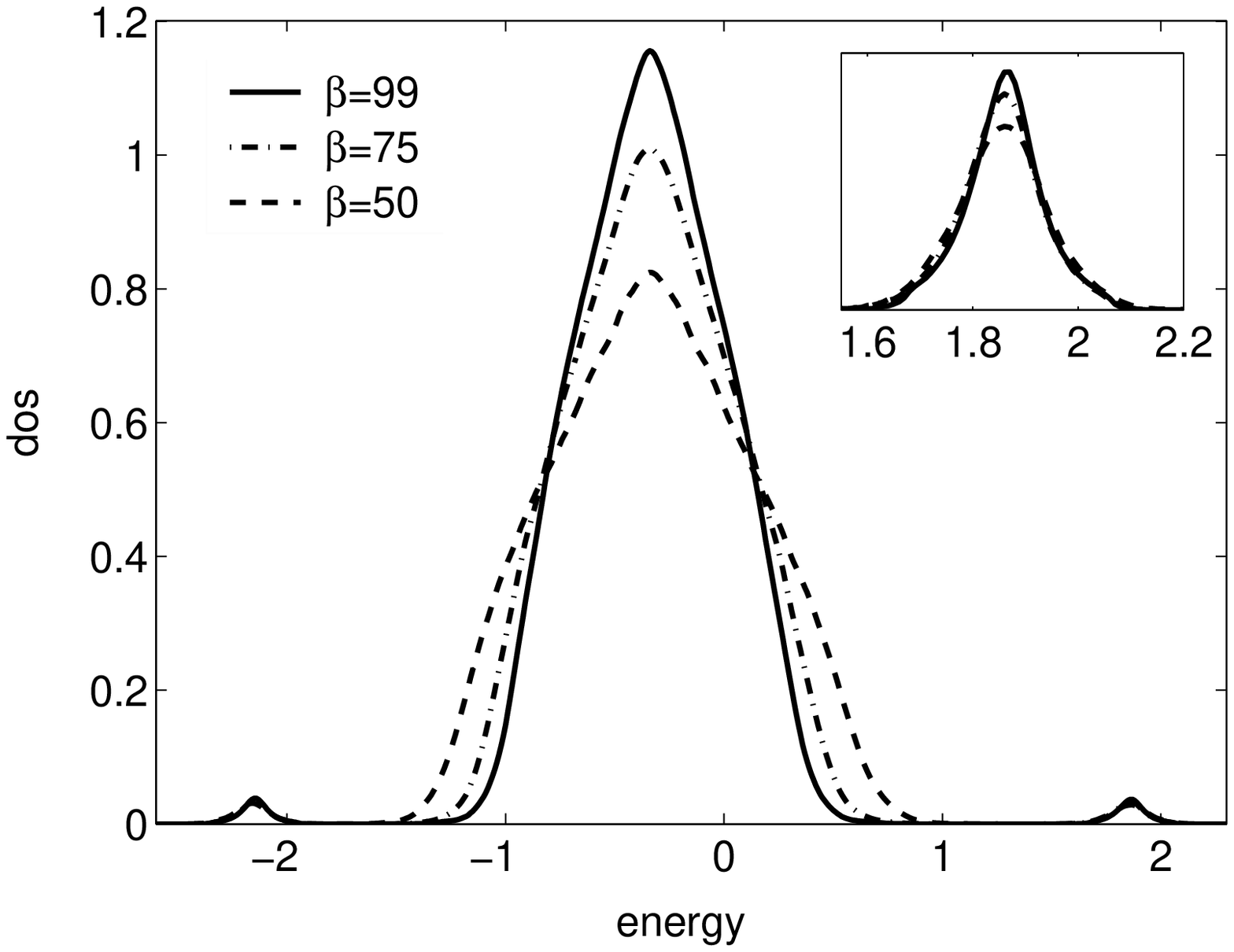}  
    \label{dos_1h_JH06J002betaX.eps}}
  \subfigure[]{\includegraphics[width=0.37\textwidth]{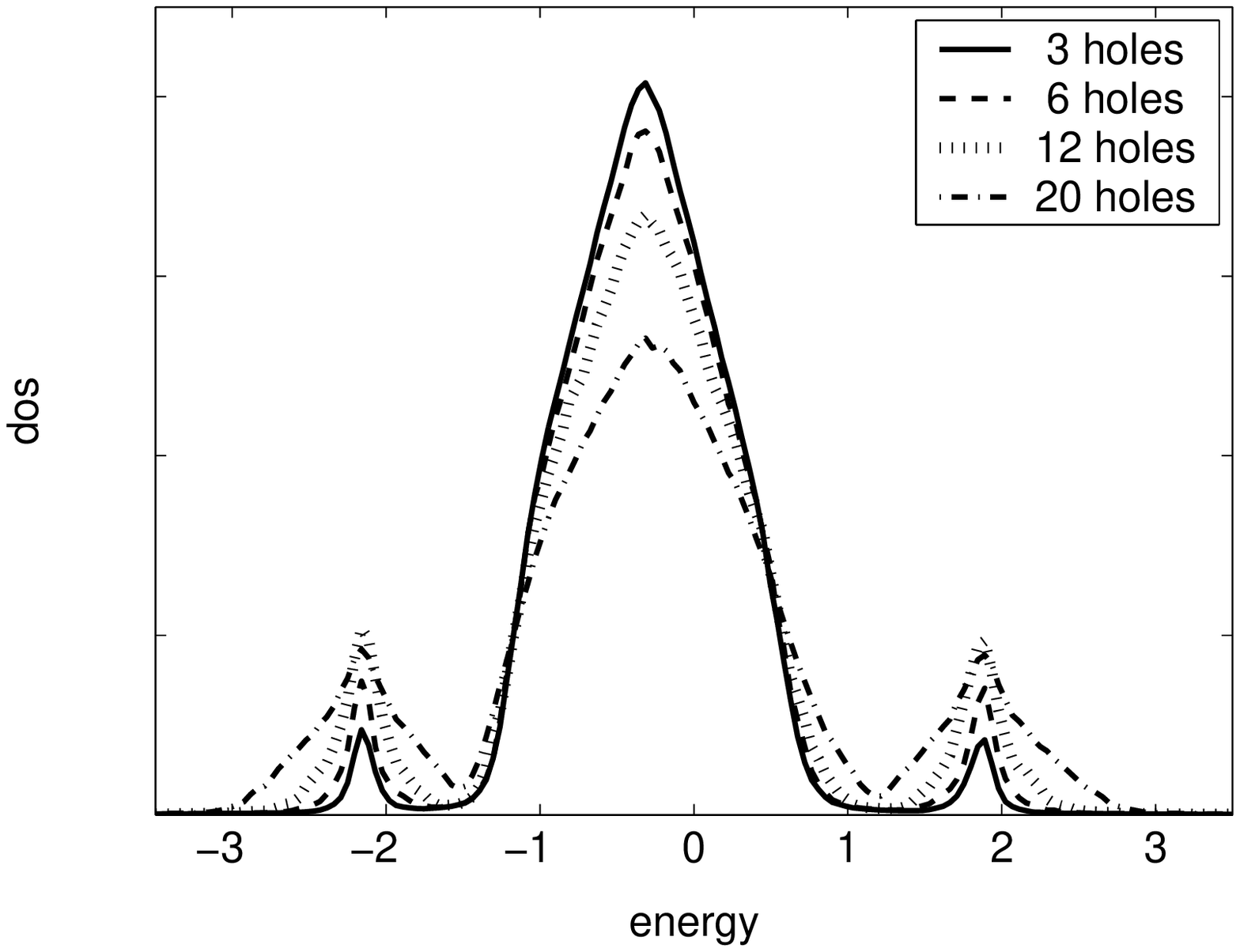}
    \label{dos_Jse0.02_beta50_Nvar.eps}}
  \caption{One-particle DOS for few holes in a $14\times 12$ lattice, $J'=0.02,
    \JH=6$.
    (a): Temperature-dependence of the DOS in the case of one
    hole. The inset shows an enlargement of the polaronic peak.
    (b) Doping dependence of the DOS for $\beta=50$\label{dos_Jse0.02}} 
\end{figure}

As in the one-dimensional case, the polarons induce separate states in the
one-particle density of states (DOS) depicted in
Fig.~\ref{dos_Jse0.02}, as described in Sec.~\ref{sec:FMP}.
Figure \ref{dos_1h_JH06J002betaX.eps} shows the DOS in the case of one
hole in a $14\times 12$ lattice at various temperatures.
One observes a broad peak in the center and two polaronic peaks at $\pm 2$
that are separated by a pseudogap. The pseudogap observed in the Kondo model
is thus a direct result of the FM polarons.
The broad central peak is due to holes moving in the not quite perfectly
antiferromagnetic background. As a result of 
the superexchange term in the Hamiltonian, it is centered around $\eps =
-z/(2\JH)$ with $z=4$ in the 2D square lattice.
When no hole is in the system, this peak also shows up and is then the only
feature of the DOS. The width of the peak is mainly dominated by corespin
fluctuations around the completely ordered state. Consequently the width decreases with
decreasing temperature. This leads to a wider pseudogap at lower temperatures,
as depicted in Fig.~\ref{dos_1h_JH06J002betaX.eps}.

The polaronic peak, on the other hand, remains largely unaffected by
temperature. As can be seen in the inset of
Fig.~\ref{dos_1h_JH06J002betaX.eps}, its shape is virtually constant.
Upon introducing more holes (see Fig.~\ref{dos_Jse0.02_beta50_Nvar.eps}),
the weight of the polaronic peaks increases whereas their position and
the shape of the central band remain unaffected. The weight of the
polaronic peaks corresponds to the number of holes. The shape of both 
the antiferromagnetic band and the polaronic peaks only begins to change when
a large number of holes are added, so that the probability for overlapping
polarons becomes considerable. With $5$ sites per polaron, $20$
polarons would take $100$ sites, or $60\%$ of a $14\times 12$ lattice,
therefore connected polarons and disturbances of the AF background are to be
expected. It is therfore remarkable that even with 20 holes, the
polarons seem to remain largely independent.

\begin{figure}[htbp]
  \centering
   \subfigure[]{\includegraphics[width = 0.48\textwidth,trim= 0 50 0
   50,clip]{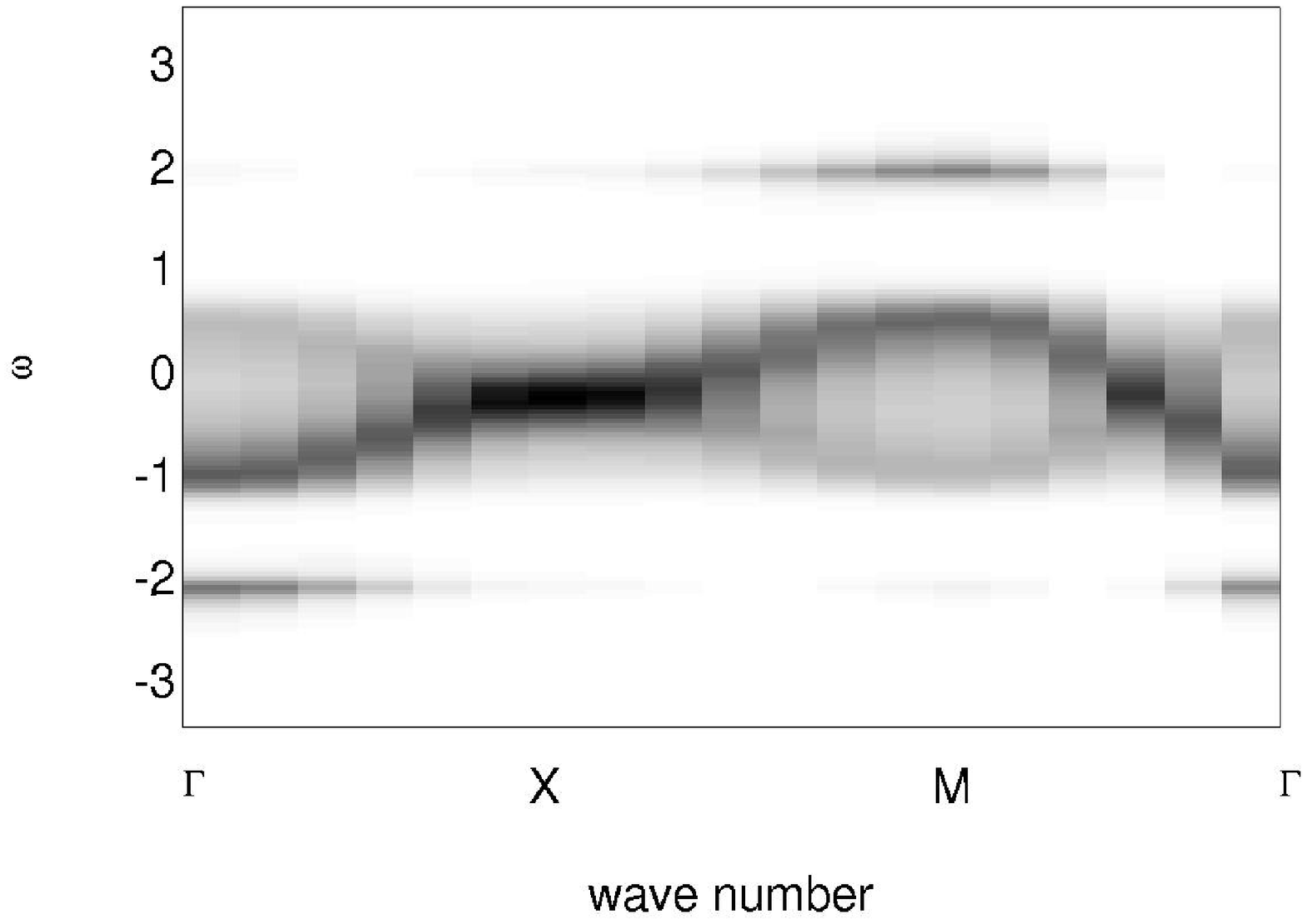}\label{fig:spec_Jse0.02_beta50_N162}}
 \subfigure[]{\includegraphics[width = 0.48\textwidth,trim= 0 50 0  50,clip]
   {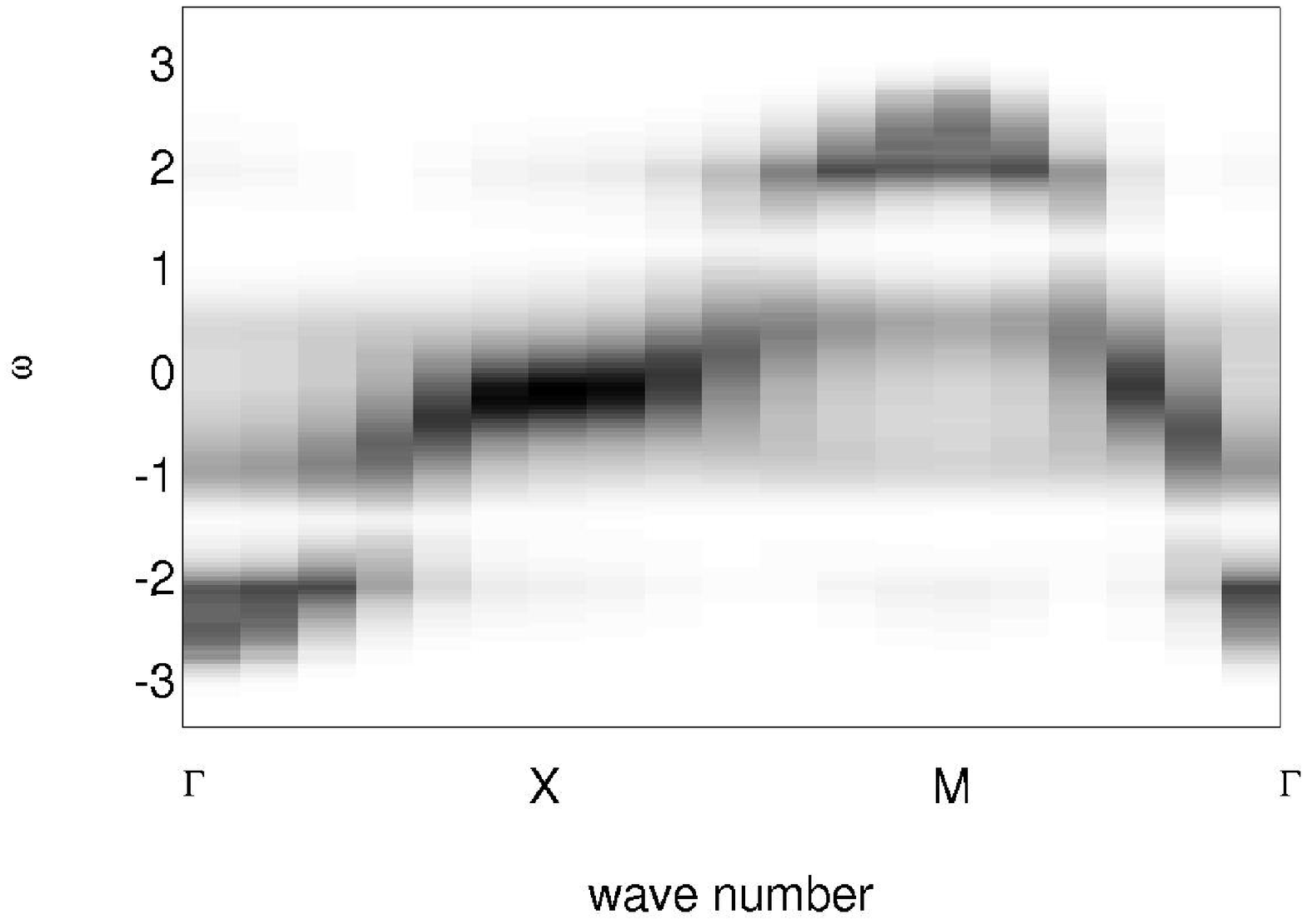}\label{fig:spec_Jse0.02_beta50_N148}}
   \caption{Spectral density for $J'=0.02, \beta=50$, $\JH=6$ on a
   $12 \times 14$ lattice. (a):
   six holes ($x\approx 0.035$): Polaronic states in addition to the AFM
     band. (b): 20 holes ($x\approx 12\%$)}
  \label{fig:spec_Jse0.02_beta50}
\end{figure}

Figure~\ref{fig:spec_Jse0.02_beta50} shows the spectral density for the
ESF model with $\JH=6$, $J'=0.02$ and $\beta=50$ for $6$ and
$20$ holes. 
In addition to the central band, the
polaronic states can be seen at energies
slightly below $\pm 2$, in perfect agreement with the simple
polaron model, see Fig.~\ref{polaron_bands}. The states at $\omega\approx
0$ are lost within the AFM band.
The results for $20$ holes
($x\approx0.12$, Fig.~\ref{fig:spec_Jse0.02_beta50_N148}) are slightly
smeared compared to the results for the simple polaron model in
Fig.~\ref{polaron20_bands.eps}, but the similarities are striking.

\begin{figure}[htbp]
  \centering
   \includegraphics[width = 0.6\textwidth]{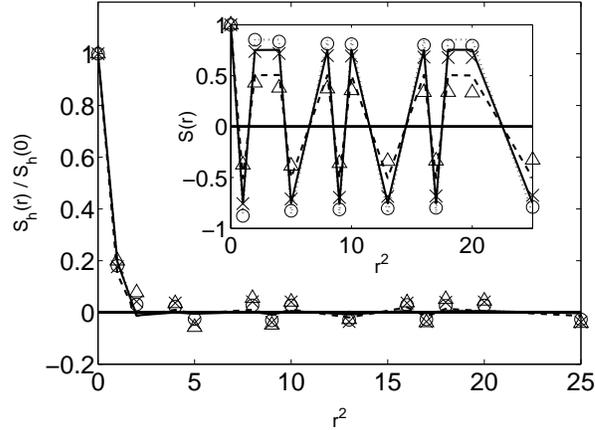}
   \caption{Dressed corespin correlation \Eq{eq:hss_def} for $J'=0.02$
     from unbiased MC Data for 1 ($\circ$), 6 ($\times$) and 20
     ($\triangle$) holes.
     Continuous lines are data for the simple Polaron model (see
     Sec. \ref{sec:FMP}): 1 Polaron (dotted), 6 (solid)
     and 20 Polarons (dashed). The inset shows the corespin correlation
     $S(\vec r) = \frac{1}{L} \sum_{\vec i} S_{\vec i} \cdot S_{\vec i + \vec r}$.
     The MC Simulations were done for $\JH=6$, $\beta=50$ on a $12 \times 14$ lattice.}
   \label{fig:hss_MCvsPolMod}
\end{figure}

In order to verify that the addition of holes leads primarily to more small
polarons rather than to a growth of the existing ones, we use a
dressed corespin correlation function 
\begin{equation}                                       \label{eq:hss_def}
  S_h(\vec r) = \frac{1}{L}
  \sum_{\vec i} n^h_{\vec i} S_{\vec i} \cdot S_{\vec i + \vec r}\;.
\end{equation}
Where $n_{\vec i}^h$ is the hole density at site $\vec i$ related to the
electron density via $n_{\vec i}^h=1-n_{\vec i}$, the sum over $\vec
i$ is taken over all lattice sites.
This dressed correlation measures the ferromagnetic regions around
holes.
Figure~\ref{fig:hss_MCvsPolMod} shows the results for $1$, $6$ and
$20$ holes on $12\times14$ sites, which correspond to doping levels of
$x=0.006$, $x=0.036$, and $x=0.12$. The results are almost independent of the
doping level $x$ and, above all, the ferromagnetic region does not grow with
increasing hole density. The data are compared to those obtained for the
independent polaron model introduced in Sec.~\ref{sec:FMP} and show very good
agreement.
The inset of Fig.~\ref{fig:hss_MCvsPolMod} shows the usual corespin
correlation which reveals the AFM background. The antiferromagnetism
decreases with increasing hole concentration, 
both for the MC simulations and the idealised polaron model.
However, it shrinks somewhat faster for the full ESF model.

The results begin to deviate from the independent polaron results only for
doping levels $x > 13 \%$, with a more homogenous phase setting in at
$x\approx 21\%$. In between, the polarons attract each other with a tendency to form
larger FM clusters in the antiferromagnetic background, which eventually leads to phase
separation (PS). 
The transition from polarons to PS is not well defined.
On increasing hole concentration the polarons first coexist with 
larger hole-rich clusters which then grow and finally dominate.
Details on this range of doping and on the influence of the AFM superexchange
parameter $J'$
are subject of a subsequent publication.

\section{Conclusions}                           \label{sec:conclusion}

In this paper, the ferromagnetic Kondo (double-exchange) model in 2D has been
analysed by unbiased finite temperature Monte-Carlo simulations.
It has been found that upon hole doping, small ferromagnetic regions appear around
each individual hole while the rest of the lattice stays antiferromagnetically
ordered. Each of the ferromagnetic regions contains one single hole.
Therefore, the physics close to half filling is not governed by phase
separation into larger FM and AF regions, as previously reported, but
by single-hole ferromagnetic polarons moving in an antiferromagnetic background.

The critical chemical potential~$\mu^*$ at which holes start to enter the lower
Kondo band can be found from simplified energy considerations (\Eq{eq:e_pol}).
For $\mu$ significantly above $\mu^*$, the band is completely filled
and the corespins are antiferromagnetic.
Around $\mu^*$, holes enter the $e_g$-band, forming isolated FM domains in
the shape depicted in Fig.~\ref{polaron_conf}, each containing one {\em
single} hole. This is corroborated by MC snapshots, the functional dependence
of the electron density on the chemical potential, the spectral density and
the dressed corespin correlation \Eq{eq:hss_def}.

The discontinuity in the electron density vs. the chemical potential (i.~e.
infinite compressibility) is usually taken as evidence for PS.
In the case of the Kondo model, this discontinuity is a consequence of a large
(macroscopic) number of degenerate polaron states.
When the chemical potential is close to the energy of these states,
the number of holes (polarons) in the lattice strongly fluctuates.
The weight of the polaron peak in the spectrum is directly linked to the
number of holes (Fig.~\ref{dos_Jse0.02_beta50_Nvar.eps}).
In order to obtain numerical results at a fixed hole number, it was necessary to
develop a canonical algorithm for our Monte Carlo simulations.

Another consequence of the formation of single-hole FM polarons is the opening of
a pseudogap.
The small FM regions of the polarons contain only a few electronic states that are
energetically well separated from each other.
Moreover, the width of the antiferromagnetic band is much smaller than the
difference between the highest and the lowest polaron states.
Therefore, no states can be found for energies between the upper edge of the
antiferromagnetic band and the highest state within the polaron. This gives rise to
a pseudogap in the one-particle spectral function.
The same arguments explain the appearance of a mirror gap well below the chemical
potential.

A pseudogap is indeed observed in experiments~\cite{DessauI, DessauII,
DessauIII} and in MC simulations for the
Kondo model~\cite{dagotto01:review,KollerPruell2002a}.
Experiments at low doped La$_{1-x}$Ca$_x$MnO$_3$ showed evidence of small FM
droplets in an AFM background~\cite{Biotteau_01,Hennion_98}.

Our analysis yields compelling evidence against the PS scenario and in favour
of FM polarons for small doping in 2D for realistic parameter values for
manganites.
A similar behaviour has been previously found for 1D.
Furthermore, the coupling to lattice degrees of freedom will additionally localise
holes and inhibit the formation of a ferromagnetic phase
(Jahn-Teller polarons)~\cite{EdwardsI,millis96:polaronsI,GBB04pre}.
It should be noted that, depending on the value of the hopping parameter $t_0$, the
temperature investigated in this paper ($\beta=50$) is in
a range of $50K - 100K$, which is in agreement with
temperatures in experiments.

%
\ack

This work has been supported by the Austrian Science Fund (FWF), project
no.\ P15834-PHY. We wish to thank the EPSRC (Grant GR/S18571/01) for financial
support.

\section*{References}


\end{document}